\documentclass{PoS}

\title{Charmed-Bottom mesons from Lattice QCD}

\ShortTitle{Charmed-Bottom mesons}

\author{\speaker{Nilmani Mathur}\thanks{Indian Lattice Gauge Theory Initiative}\\
        Department of Theoretical Physics, Tata Institute of Fundamental Research, Homi Bhabha Road, Mumbai 400005, India\\
        E-mail: \email{nilmani@theory.tifr.res.in}}

\author{M. Padmanath\\
        Institut  fur  Theoretische  Physik,  Universitat  Regensburg,
Universitatsstrase  31,  93053  Regensburg,  Germany\\
        E-mail: \email{Padmanath.M@physik.uni-regensburg.de}}

\author{Randy Lewis\\
        Department of Physics \& Astronomy, York University, Toronto, ON M3J 1P3, Canada\\
        E-mail: \email{randy.lewis@yorku.ca}}

\abstract{We present ground state spectra of mesons containing a charm and a bottom quark. For the charm quark we use overlap valence quarks while a non-relativistic formulation is utilized for the bottom quark on a background of 2+1+1 flavors HISQ gauge configurations generated by the MILC collaboration. The hyperfine splitting between $1S$ states of $B_c$ mesons is found to be $56^{+4}_{-3}$ MeV. We also study the baryons containing only charm and bottom quarks and predict their ground state masses. Results are obtained at three lattice spacings.}

\FullConference{34th annual International Symposium on Lattice Field Theory\\
		24-30 July 2016\\
		University of Southampton, UK}

\begin{document}

\section{Introduction}
Lattice QCD methods provide a unique opportunity to study hadronic
physics, particularly the energy spectra of hadrons. Substantial
progress has been made to extract the ground and the excited states of
charmed hadrons, particularly for charmonia.  However, the study of
bottom hadrons with relativistic actions and controlled
discretizations is still prohibitively computer intensive though
recent progress in relativistic heavy quark actions is promising. Most
studies involving bottom quarks are based on the non-relativistic QCD
(NRQCD) formulation, heavy quark effective theory (HQET) and the static
quark formulation.

Study of heavy mesons plays a very important role in understanding the nature
of strong interactions. Though heavy quarkonia and heavy light mesons
have been investigated in great detail, not much is known about the
heavy mesons containing only charm and bottom quarks. It is expected
that the physics of charmed-bottom mesons involves multiple scales :
$1/m_b$ ($v_b = 0.05$), $1/m_c$ ($v_c = 0.4-0.5$), and $\Lambda_{QCD}$. 
It is interesting to investigate whether charmed-bottom mesons 
behave like heavy-light mesons or like quarkonia states. 
The information about hyperfine splittings and other spin
splittings, in addition to the results on decay constants, can shed light
on the structure of these states. 

Experimentally only one state, $B_c(0^{-})$, is established with mass at 6275(1)
MeV~\cite{PDG}. Recently ATLAS observed another $B_c$ meson with mass
$6842\pm 4 \pm 5$ MeV, which was interpreted as the excited state,
$B_c^{0^{-}}(2S)$~\cite{Aad:2014laa}. However, this excitation has not 
been confirmed by
other experiments yet. On the theoretical side, potential model
predictions for these states vary widely. For example, the prediction
for $1S$ hyperfine splitting varies in the range 40-90 MeV~\cite{q1,q2,Kiselev:1994rc,Ebert:2002pp,Godfrey:2004ya}. This is due to various 
ways of tuning the heavy quark potentials with spin dependent 
terms and less clarity 
of the wavefunction at the origin~\cite{q1,q2,Kiselev:1994rc,Ebert:2002pp,Godfrey:2004ya}. On the other hand, using lattice QCD
methods, though heavy-light or heavy-heavy mesons were studied
extensively not many calculations were carried out for charmed-bottom
mesons and baryons. Only two collaborations, HPQCD~\cite{Gregory:2009hq, Dowdall:2012ab} and Wurtz {\it et al.}~\cite{Wurtz:2015mqa}, have studied charmed-bottom mesons recently. Similarly, there are very few recent results for charmed-bottom baryons~\cite{Francis:2016,Brown:2014ena}.

In this report, we present our preliminary results on charmed-bottom
mesons and baryons. For the bottom quark we use an NRQCD action
with non-perturbatively tuned coefficients with terms up to
$\mathcal{O}(v^4)$, whereas the overlap action is utilized for the valence
charm quark. Hyperfine splitting between $1S$ $B_c$ states as well as
masses for other $B_c$ mesons are predicted. Moreover, ground state
energy spectra of baryons with charm and bottom quarks are also
predicted.

\vspace*{-0.1in}

\section{Simulation Details}

\vspace*{-0.1in}

We use three sets of dynamical 2+1+1 flavours HISQ gauge field ensambles
generated by the MILC collaboration : $24^3 \times 64$, $32^3 \times
96$ and $48^3 \times 96$ lattices at gauge couplings $10/g^2 = 6.00,
6.30$ and $6.72$, respectively. The details of these gauge
configurations are summarized in Ref.~\cite{Bazavov:2012xda}. We use
the unphysical $\bar{s} s$ pseudoscalar mass equal to 685 MeV to tune
the strange quark mass while the $\Omega_{sss}$ baryon mass is used for
calculating the lattice spacings as mentioned in
Refs.~\cite{Basak:2012py,Basak:2013oya}.  The measured lattice
spacings, 0.1192(14), 0.0877(10) and 0.0582(5) {\it{fm}}, are
consistent with 0.1207(11) 0.0888(8) and 0.0582(5) {\it{fm}}, respectively,
measured by MILC collaboration with this set of ensembles using the
$r_1$ parameter.

We have adopted an NRQCD formulation for the bottom quark.
The non-relativistic action that we use is discussed in Ref.~\cite{Lepage:1992tx}. We have considered all terms up to $1/M_{0}^2$ and the leading term of the order of $1/M_{0}^3$, where $M_{0} = am_b $ is the bare mass for bottom quark in lattice units. The NRQCD Hamiltonian is given by,  
\begin{equation}
H = H_{0} + \delta H,
\end{equation}
where $H_{0}$ is the kinetic term and is defined by,
\begin{equation}
H_{0} = - {{\Delta^{(2)}}\over{2M_0}},
\end{equation}
whereas, $\Delta H$ contains interaction terms and is given by,
\begin{eqnarray}
\Delta H &=& - c_1 {{(\Delta^{(2)})^2}\over{8M_0^3}} + 
c_2 {{i}\over {8M_0^2}} {\left(\tilde{\nabla} \cdot {\bf\tilde{E}} - {\bf\tilde{E}} \cdot \tilde{\nabla}\right)}
- c_3 {{1}\over{8M_0^2}} \bf{\sigma} \cdot {\left(\tilde{\nabla} \times {\tilde{\bf{E}}} - {\tilde{\bf{E}}} \times \tilde{\nabla}\right)} \nonumber \\
&& - c_4 {{1}\over{2M_0}} {\bf{\sigma}} \cdot {\tilde{\bf{B}}} \,+  c_5 {{\Delta^{(4)}}\over{24M_0}} \, -  c_6 {{(\Delta^{(2)})^2}\over{8nM_0^2}} .
\end{eqnarray}
The quantities with a tilde assume an $\mathcal{O}(a)$ corrected form of discretization. $\nabla$ is the symmetric lattice derivative while $\Delta^{(2)}$ and $\Delta^{(4)}$ represent the lattice discretized versions of $\sum_{i}D^2$ and $\sum_{i}D^4$, respectively.  More details of this Hamiltonian can be found in Ref.~\cite{Lewis:2008fu}. This Hamiltonian is improved by including spin-independent terms through $\mathcal{O}(v^4)$. For the coarser two ensembles, we use the values of the improvement
coefficients, $c_1$ to $c_6$, as estimated non-perturbatively by the HPQCD collaboration~\cite{Dowdall:2011wh}. For the finer lattice, we use tree level coefficients. The NRQCD quark propagators are obtained by usual time evolution
%\begin{eqnarray}
%G({\bf{x}},t+1) = &&\left(1 - {\delta H\over 2}\right)\left(1 - {\delta H_0\over 2n}\right)^nU^{\dagger}_t(x)\\
%&& \quad \times \left(1 - {\delta H_0\over 2n}\right)^n\left(1 - {\delta H\over 2}\right)  G({\bf{x}},t).
%\end{eqnarray}
\begin{equation}
G({\bf{x}},t+1) = \left(1 - {\delta H\over 2}\right)\left(1 - {\delta H_0\over 2n}\right)^nU^{\dagger}_t(x)
 \left(1 - {\delta H_0\over 2n}\right)^n\left(1 - {\delta H\over 2}\right)  G({\bf{x}},t).
\end{equation}
A wall source is utilized as smearing function for calculating these quark propagators. 

The bottom quark mass is tuned by equating the lattice spin-average mass of $1S$ bottomonium to its experimental value. The lattice spin-average mass is obtained from the kinetic mass relation :
\begin{equation}
\bar{M}_{kin}(1S) = {3\over 4} aM_{kin}(\Upsilon) + {1\over 4} aM_{kin}(\eta_b),
\end{equation}
where the kinetic mass is calculated from the relativistic energy-momentum dispersion relation :
\begin{equation}
aM_{kin} = {{a^2p^2 - (a\Delta E)^2}\over{2a\Delta E}}.
\end{equation}
Here $a\Delta E$s are extracted from the energy difference between the
mesons with momenta $pa$ and zero.  A momentum induced wall-source is
utilized to obtain energy values from the correlators with finite
momenta. This method was found to be very efficient compared to point
or smeared sources~\cite{Basak:2012py} and helps to obtain kinetic
masses precisely with significantly little statistics.

For valence charm quark propagators we use the overlap action as
described in Refs.~\cite{Basak:2012py,Basak:2013oya}. The overlap
action does not have $\mathcal{O}(ma)$ errors, and it is chirally
symmetric at finite lattice spacings. The charm mass is tuned by
equating the spin-averaged kinetic mass of the $1S$ charmonia states to its
physical value. The tuned bare charm quark masses ($am_c$) are found
to be 0.528, 0.425 and 0.29 on coarser to finer lattices
respectively. Details of the charm tuning was described in Ref.~\cite{Basak:2013oya}.

\section{Results}
\vspace*{-0.15in}
\subsection{Charmed-bottom mesons}
\vspace*{-0.05in}
Lattice methods suffer from discretization errors, particularly for
hadrons with heavy quarks. A good agreement of hyperfine splitting of
the $1S$ mesons on the lattice with the respective experimental value
ensures good control over the discretization errors and hence a
reliable estimation of the heavy meson spectra using lattice
methods.  Figure 1 shows our results for the hyperfine splittings in
$1S$ quarkonium for three different ensembles we use. Figure 1(a) is for
$1S$ charmonium showing our results with filled circles and experimental
value with star symbol. An extrapolation with a form $1/a^3$, including systematics, gives a fit value 115(3) MeV for this splitting which is consistent with the
experimental value 113.9(6) MeV. We have excluded the disconnected
diagrams which is shown to reduce the splittings by a few MeV~\cite{Levkova:2010ft}. Figure 1(b) is for $1S$ bottomonium. Black circles are 
results from tree-level
coefficients and blue are with improved coefficients in the NRQCD
action (this color coding will be followed throughout). A naive fit combining the results from improved coefficients
for coarser lattices and the result from unimproved coefficients at
the finer lattice gives $1S$ bottomonium hyperfine splitting as 64(3) MeV,
whereas the experimental value is $62.3\pm 3$ MeV. These results show
that the discretization errors in our calculation for quarkonia are
well within control.
%0.35 0.325
\begin{figure}[h]
\vspace{-0.1in}
%\begin{center}
\parbox{.45\linewidth}{
\centering
\includegraphics[width=0.5\textwidth,height=0.35\textwidth,clip=true]{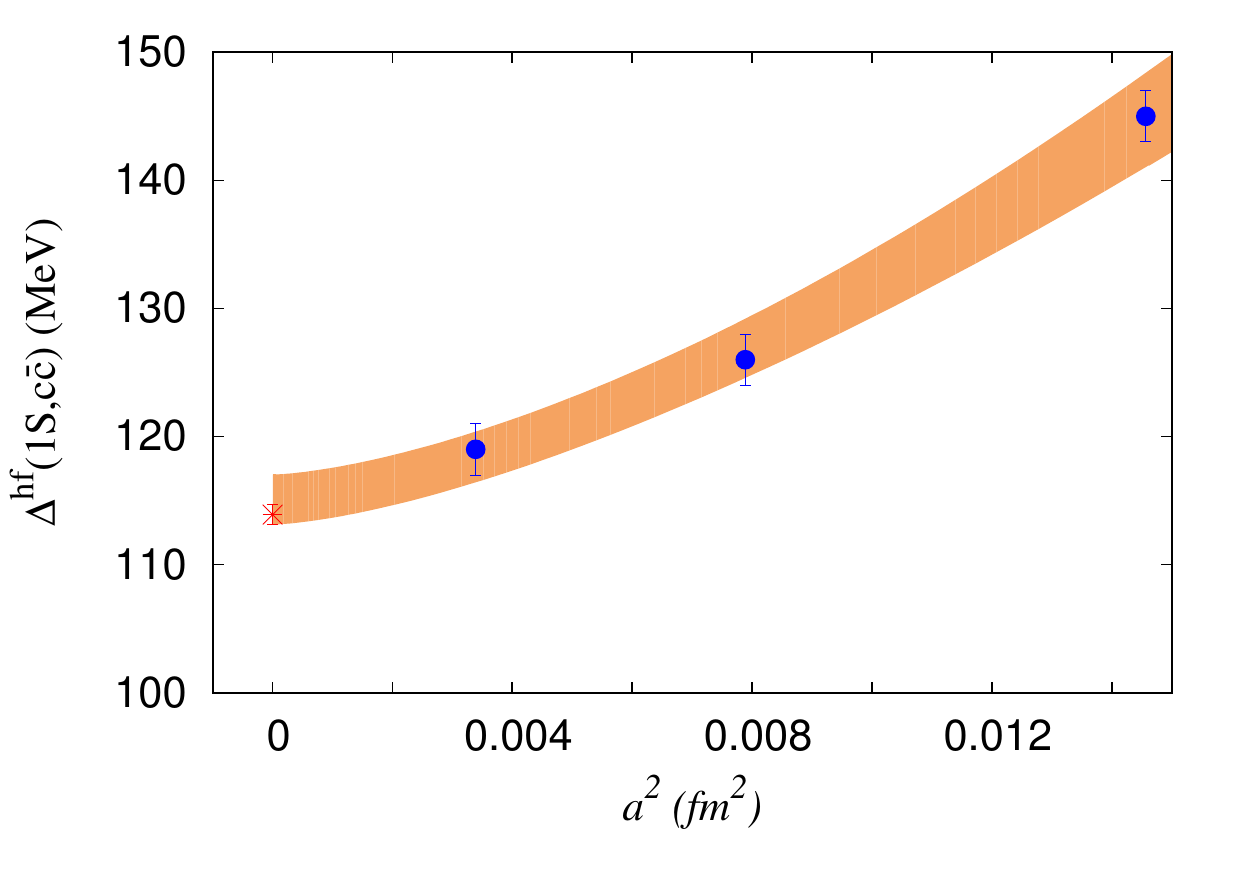}\\
(a)}
\hspace{0.75cm}
\parbox{.45\linewidth}{
 \centering
\includegraphics[width=0.45\textwidth,height=0.325\textwidth,clip=true]{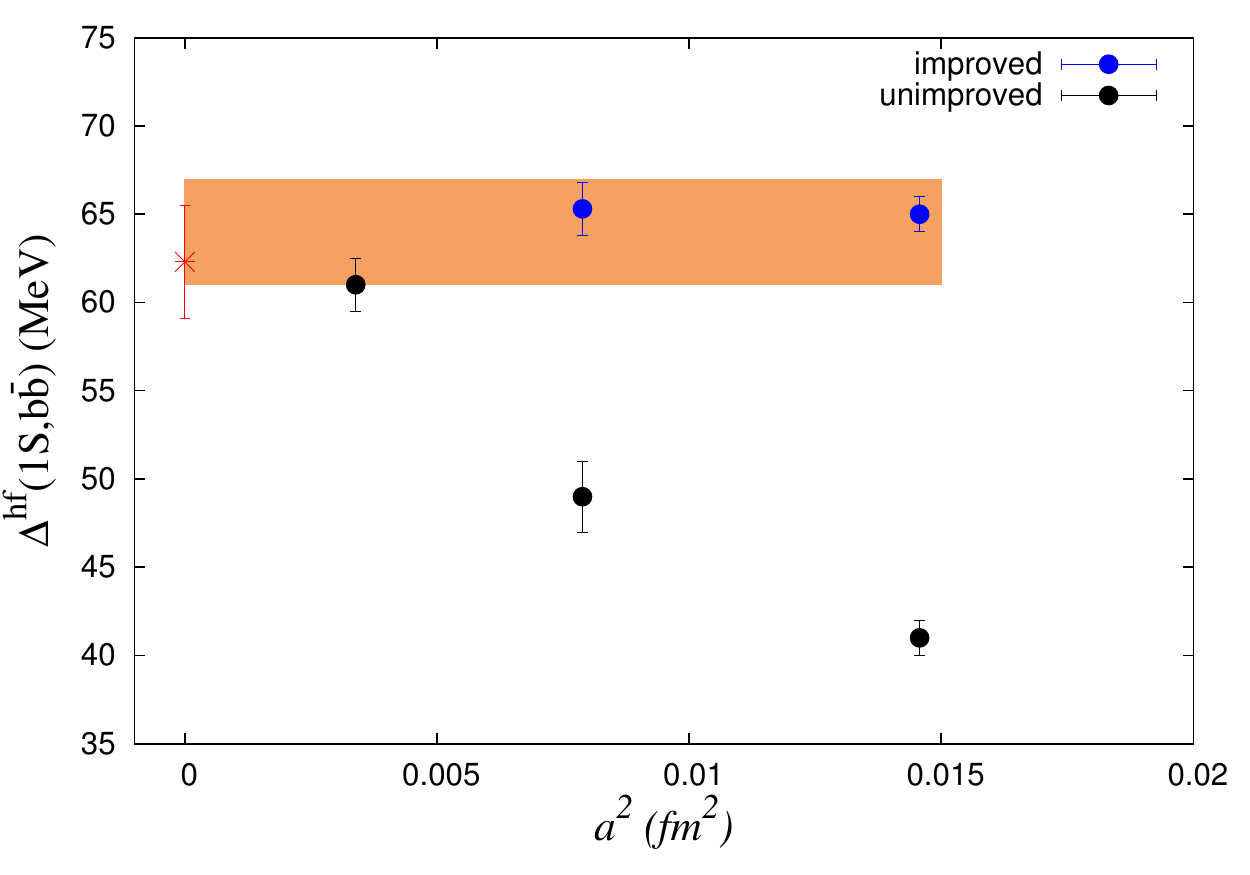}\\
(b)}
%\end{center}
\vspace{-0.1in}
\caption{Hyperfine splittings of $1S$ quarkonium plotted against the square of the lattice spacing for (a) charmonium and (b) bottomonium at three lattice spacings.
See explanation in the text for symbols.
\label{fig_hyp_cc_bb}}
\end{figure}

\vspace*{-0.1in}

\begin{figure}[h]
\vspace{-0.1in}
%\begin{center}
\parbox{.45\linewidth}{
\centering
\includegraphics[width=0.45\textwidth,height=0.325\textwidth,clip=true]{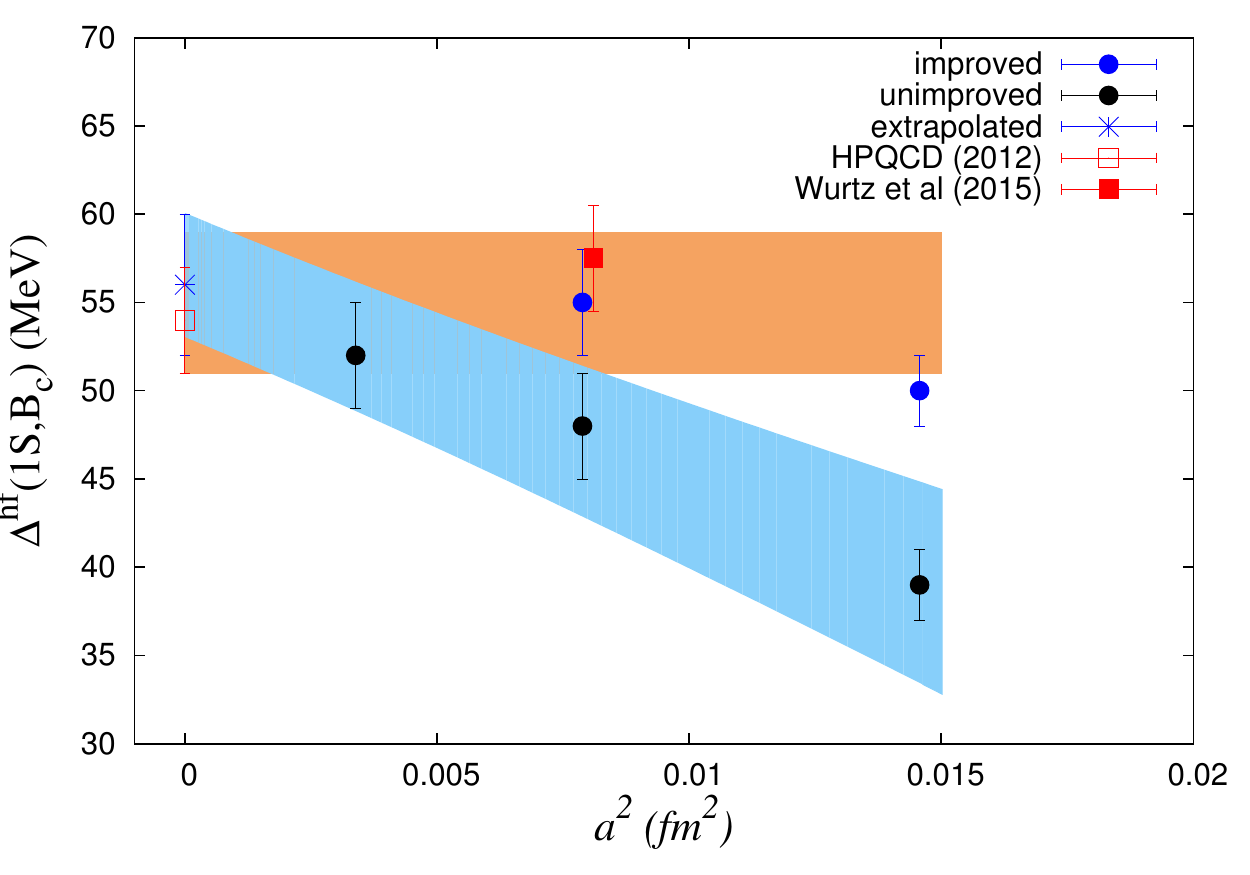}\\
(a)}
\hspace{0.75cm}
\parbox{.45\linewidth}{
 \centering
\includegraphics[width=0.45\textwidth,height=0.325\textwidth,clip=true]{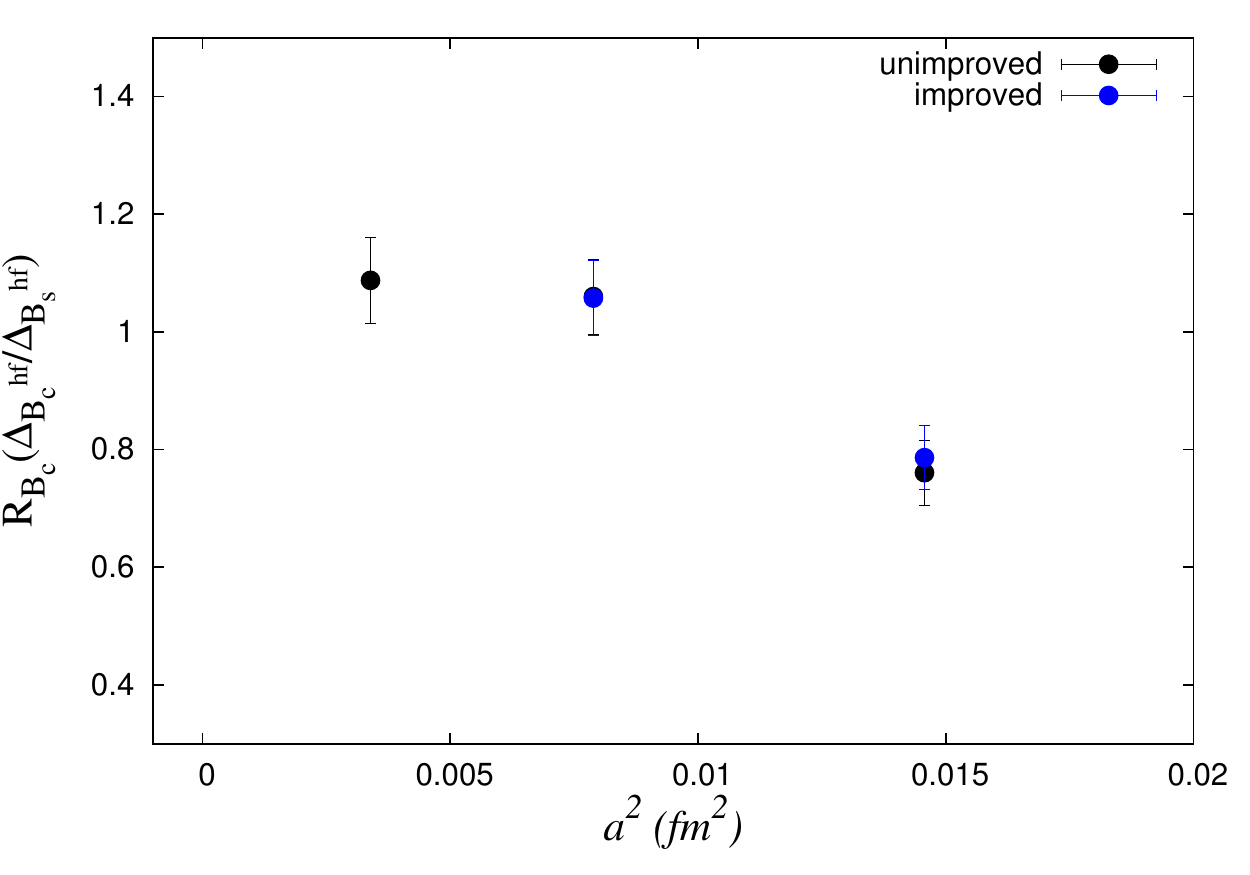}\\
(b)}
%\end{center}
\vspace{-0.1in}
\caption{(a) Hyperfine splitting of $1S$ energy levels of $B_c$ mesons at three lattice spacings. (b) The ratio of hyperfine splittings for $B_c$ to $B_s$ mesons.
\label{fig_hyp_bc}}
\end{figure}

In figure 2(a), we show our main results on the hyperfine splittings of $B_c$ mesons : $M_{B^*_c}-M_{B_c}$.
%In figure 2(a), we show this mass splitting is s
%with the same color-symbol conventions as in figure 1(b). 
Results from other lattice calculations are shown by red squares.
A combined fit with results from improved and tree-level coefficients yields an estimate for this
splitting as $56^{+4}_{-3}$ MeV (shown by blue star) which is consistent with predictions from other lattice
calculations~\cite{Dowdall:2012ab,Wurtz:2015mqa}. With an NRQCD action the hyperfine splitting of quarkonia is
proportional to the $c^{2}_4$ term, which is $\mathcal{O}(\alpha^2_{s})$
in our case, and it also depends on higher order ($\mathcal{O}(v^6)$)
operators. These higher order corrections would
also be present in the $B_c$ mesons, which could be reduced in the ratio of such hyperfine splittings,
e.g. between $B_s$ and $B_c$ mesons~\cite{Dowdall:2012ab}.  Similar to HPQCD~\cite{Dowdall:2012ab}  we
have constructed the following ratio,
\begin{equation}
R_{B_c} = {{\Delta^{hyp}_{B_c}}\over {\Delta^{hyp}_{B_s}}} = {{E_0(B_c^*) - E_0(B_c)}\over {E_0(B_s^*) - E_0(B_s)}}.
\end{equation}
In figure~\ref{fig_hyp_bc}(b) we show this ratio at three lattice spacings.

\vspace*{-0.3in}

\begin{figure}[h]
\vspace{-0.1in}
%\begin{center}
\parbox{.45\linewidth}{
\centering
\includegraphics[width=0.45\textwidth,height=0.3\textwidth,clip=true]{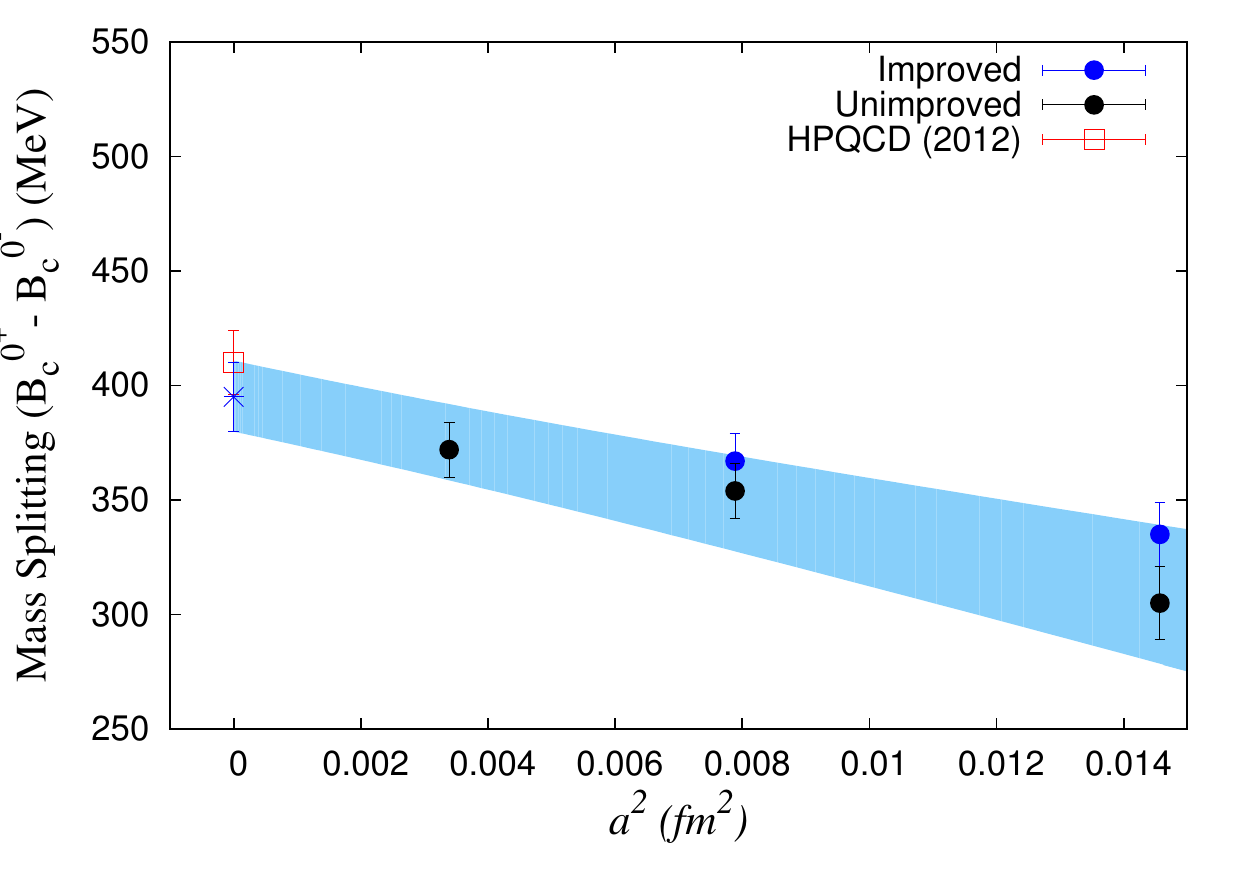}\\
(a)}
\hspace{0.75cm}
\parbox{.45\linewidth}{
 \centering
\includegraphics[width=0.45\textwidth,height=0.3\textwidth,clip=true]{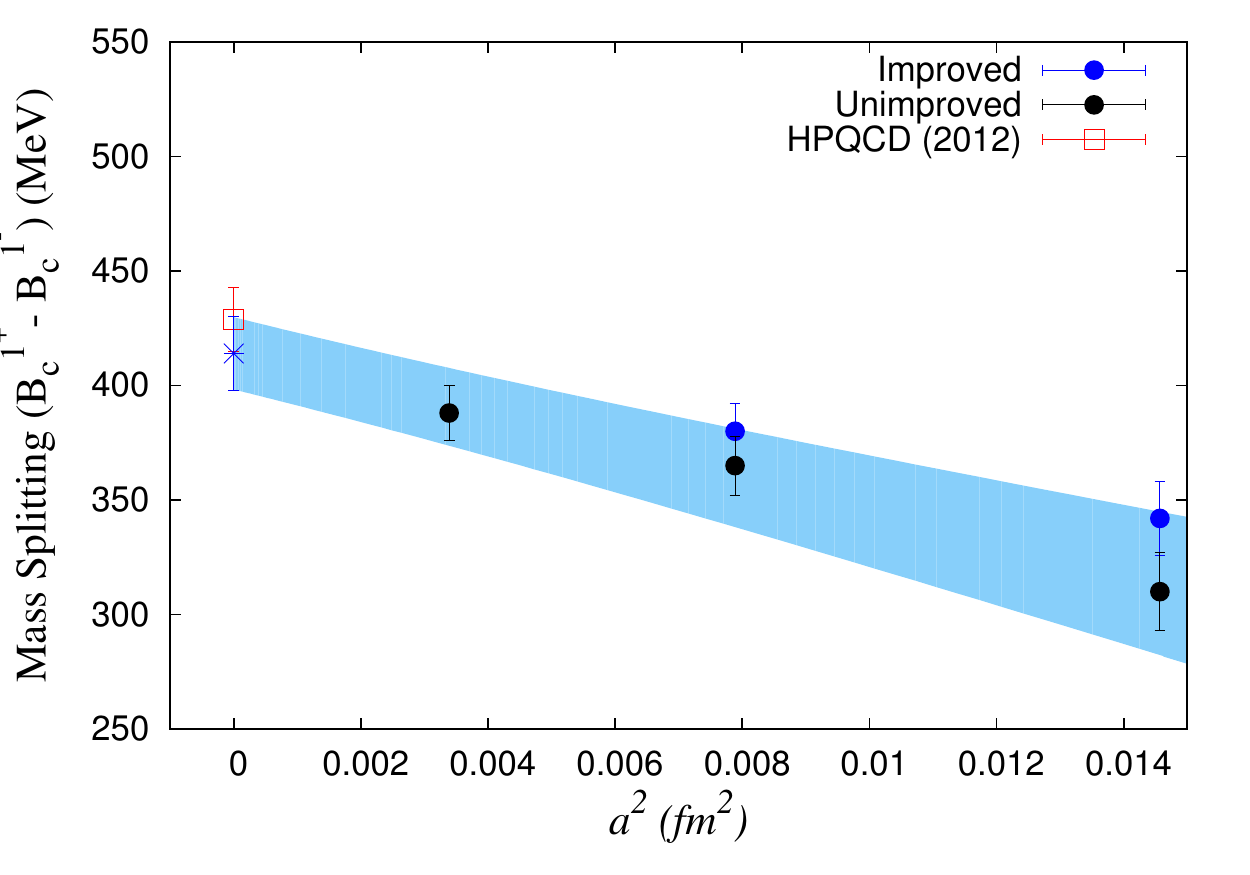}\\
(b)}
%\end{center}
\vspace{-0.1in}
\caption{Mass splittings between (a) $0^{+}-0^{-}$ and (b)  $1^{+}-1^{-}$ states in $B_c$ mesons.
\label{fig_Bc_ax_sc}}
\end{figure}

We also calculate the ground state masses of the axial-vector 
and scalar $B_c$ mesons.
In figure~\ref{fig_Bc_ax_sc}(a) we plot the splitting between scalar and
pseudoscalar states ($0^{+}-0^{-}$) while in figure 3(b) the
splitting between axial vector and vector states ($1^{+}-1^{-}$) is
shown. HPQCD results are shown with open square. After extrapolation with a form $1/a^2$ to the data from unimproved coefficients we obtain $\Delta_{B_c}^{0^{+}-0^{-}} = 414(16)$ MeV and $\Delta_{B_c}^{1^{+}-1^{-}} = 395(15)$ MeV (shown by blue star). Taking the experimental value of $B_c(0^{-})$, the ground state masses for the scalar and axial-vector $B_c$ mesons are $M_{B_c(0^{+})}$ = 6690(17) MeV and  $M_{B_c(1^{+})}$ = 6726(16) MeV respectively. 
\vspace*{-0.15in}
\subsection{Charmed-bottom baryons}
\begin{figure}[b]
\vspace{-0.1in}
%\begin{center}
\parbox{.45\linewidth}{
\centering
\includegraphics[width=0.45\textwidth,height=0.3\textwidth,clip=true]{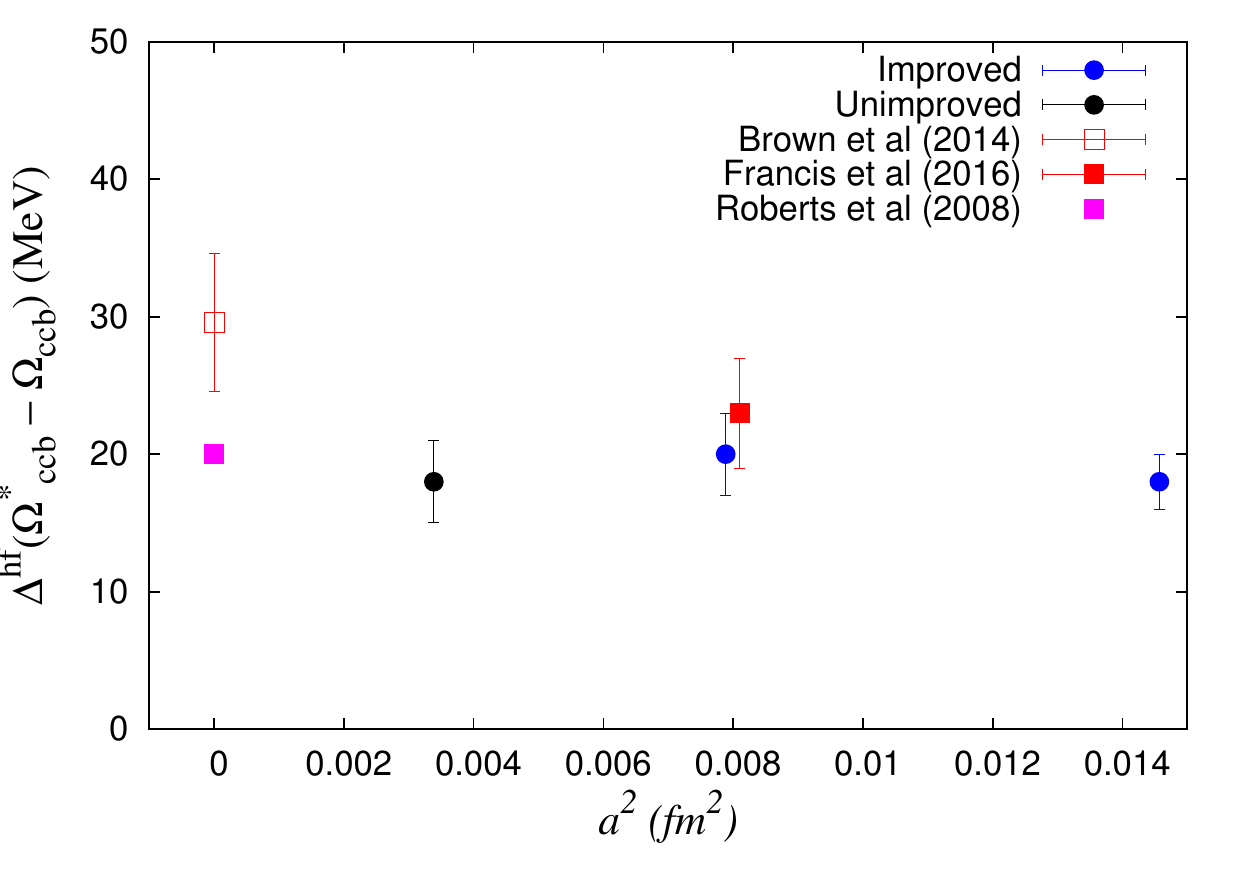}\\
(a)}
\hspace{0.75cm}
\parbox{.45\linewidth}{
 \centering
\includegraphics[width=0.45\textwidth,height=0.29\textwidth,clip=true]{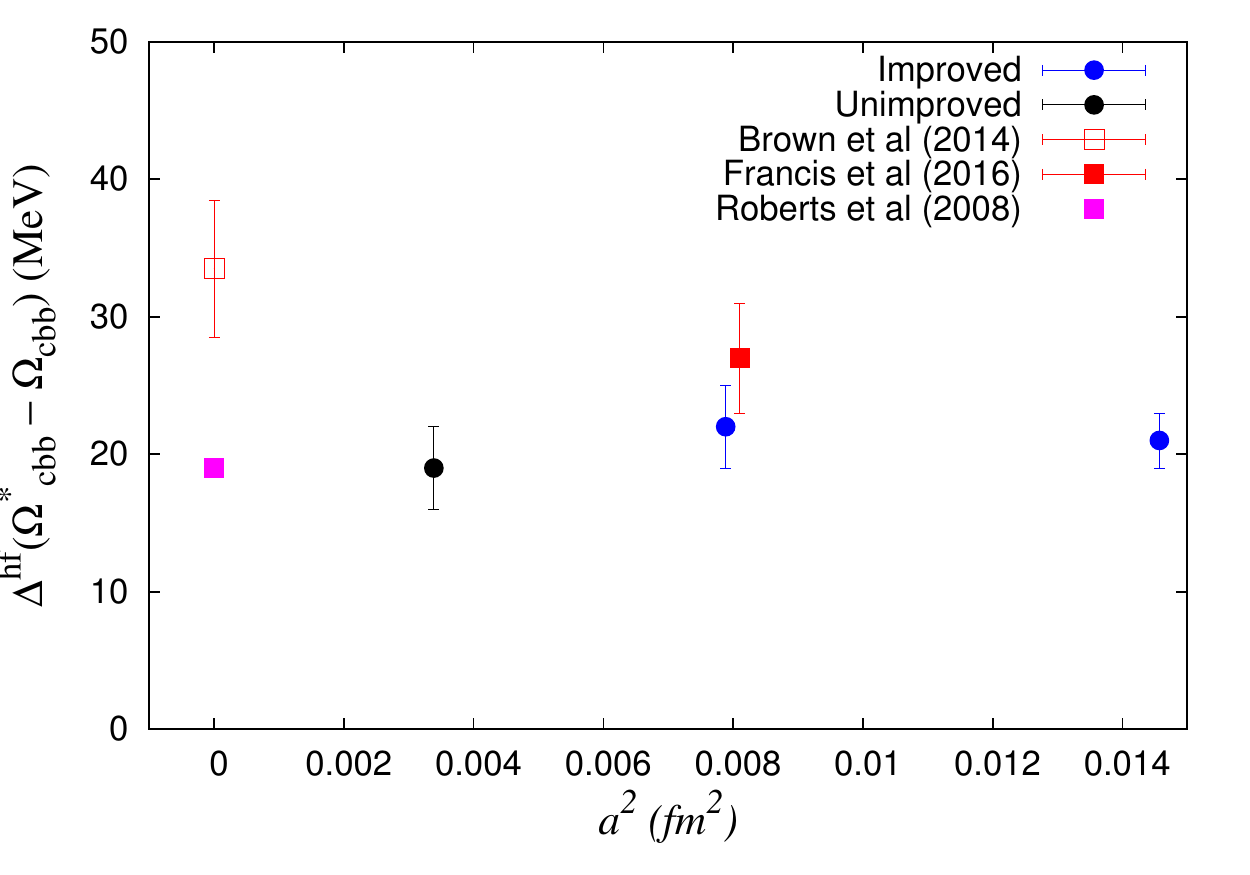}\\
(b)}
%\end{center}
\vspace{-0.1in}
\caption{Hyperfine mass splittings between (a) $\Omega^*_{ccb}-\Omega_{ccb}$ and (b)  $\Omega^*_{cbb}-\Omega_{cbb}$ baryons.
\label{fig_bc_baryons}}
\end{figure}

Baryons with three heavy quarks are interesting systems as they can
provide important information about the potential between three heavy
quarks and they can also be a study ground for effective field
theories as well as perturbative QCD. Here we present 
the ground state masses of baryons containing only charm and
bottom quarks, namely $\Omega_{ccb}({1\over 2}^+),
\Omega^*_{ccb}({3\over 2}^+), \Omega_{cbb}({1\over 2}^+)$ and
$\Omega^*_{cbb}({3\over 2}^+)$. Very similar to mesons, the hyperfine
splittings between these baryons can provide important
information about the spin dependent interactions in heavy quark
systems. In figure~\ref{fig_bc_baryons} we plot these
splittings along with other lattice
results~\cite{Francis:2016,Brown:2014ena} and quark model
predictions~\cite{Roberts:2007ni}.

\begin{figure}
\centering
\includegraphics[width=0.45\textwidth]{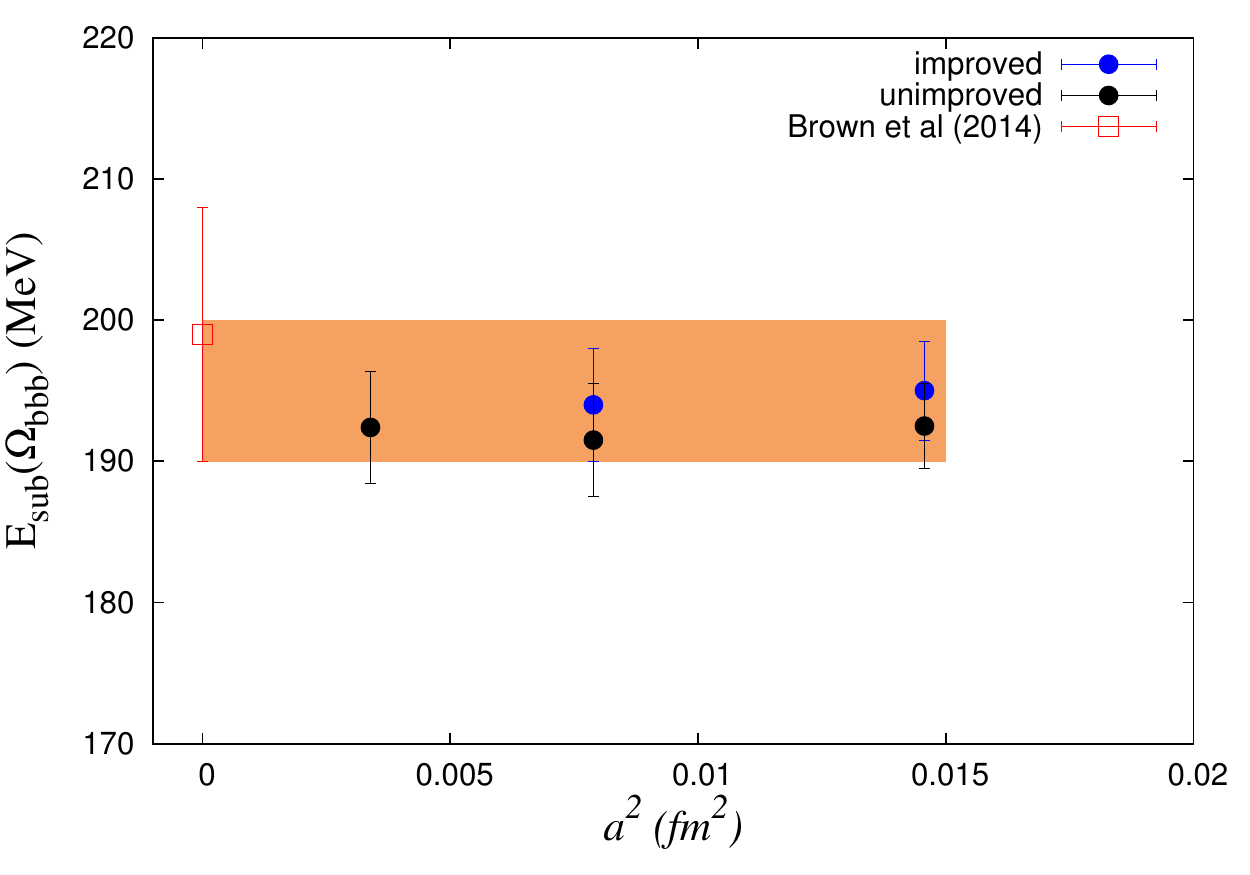}
\vspace{-0.15in}
\caption{The ground state energy of the spin-3/2 $\Omega_{bbb}$ baryon : The subtracted energy $E_{sub} = M(\Omega_{bbb}) - {3\over 2}{\bar M}(\bar bb)$, where ${\bar M}(\bar bb)$ is the $1S$ spin average bottomonium mass, is plotted at different lattice spacings.
\label{fig_triply_bottom}}
\end{figure}

Finally, we present results for the triply bottom baryon $\Omega_{bbb}({3\over
  2}^+)$, which may be viewed as baryonic analogues of
bottomonium. Being heavier with three bottom quarks it provides a good
 arena to test the discretization error in a lattice calculation with bottom
quarks. For the relative removal of discretization error due to the bottom
quark we calculate the subtracted energy defined as $E_{sub}^{\Omega_{bbb}} =
M(\Omega_{bbb}) - {3\over 2} {\bar{M}}(\bar bb)$, where ${\bar{M}}(\bar bb)$ is the
lattice estimate of the spin average $1S$ bottomonium mass. Figure~\ref{fig_triply_bottom} shows
our results for this subtracted energy at different lattice spacings
with improved and tree-level coefficients along with the result from
another lattice calculation~\cite{Brown:2014ena}. Flatness of the plot
shows that discretization errors due to bottom quarks after subtraction
are well within control. Combining data from improved coefficients, the
prediction for the ground state spin-3/2 triply bottom baryon is
following : $E^{\Omega_{bbb}}_{sub} = 194^{+4}_{-3}$ MeV, and taking the PDG spin average mass for $1S$ bottomonium as 9445(3) MeV, $M_{\Omega_{bbb}} = 14362^{+5}_{-4}$ MeV.

\vspace*{-0.1in}
\section{Conclusions}
\vspace*{-0.1in}
We report the ground state energy spectra of charmed-bottom mesons. An
NRQCD Hamiltonian with non-perturbatively improved coefficients is
used for bottom quarks while a relativistic overlap action is adopted
for charm quarks on a background of 2+1+1 flavours HISQ gauge
configurations generated by the MILC collaboration. Results are
obtained at three lattice spacings with physical volume of
about 3 {\it fm}. We tune the bottom and the charm quark masses by
equating the lattice spin-averaged kinetic masses of $1S$ states in
bottomonia and charmonia with their experimental values. The hyperfine
splittings between these $1S$ states are found to be 64(3) and $115(3)$ MeV
for bottomonia and charmonia, respectively, which are quite consistent
with their experimental values. Our result on hyperfine splitting
between $1S$ states of $B_c$ mesons is $56^{+4}_{-3}$ MeV. This result is
consistent with HPQCD~\cite{Dowdall:2012ab} as well as Wurtz {\it et al.}~\cite{Wurtz:2015mqa} who found this
splitting to be 54(3) and 57.5(3) MeV, respectively. It is interesting to
note that all lattice results are inconsistent with many quark model
results which vary in the range between 40-90 MeV~\cite{q1,q2,Kiselev:1994rc,Ebert:2002pp,Godfrey:2004ya}. Due to this
rather small splitting, detection of $B_c^{*}(1^-)$ from
its $B_c\gamma$ decay may turn out to be challenging in experiments.
Taking the PDG value of $B_c(0^-)$ state as 6275(1) MeV, our prediction 
for the ground state mass of $B^{*}_c$ is $6331^{+4}_{-3}$ MeV.
We also calculate the ground state energy of other $B_c$ mesons, 
namely, $B_c(0^+)$ and $B_c(1^+)$. Our preliminary results on the 
splittings between $(0^+ - 0^-)$ and $(1^+ - 1^-)$ states are 
found to be 411(16) MeV and 395(15) MeV, respectively.  The ground state masses of baryons containing only charm and
bottom quarks are also extracted. The hyperfine splittings between spin-3/2 and spin-1/2
$\Omega(ccb)$ and $\Omega(cbb)$ states are found to be 20(5) MeV 
and 21(5) MeV, respectively. Our prediction for the ground state spin-3/2
triply bottom baryon is $M_{\Omega_{bbb}} = 14362^{+5}_{-4}$ MeV.
\vspace*{-0.1in}
\section{Acknowledgement}
%\vspace*{-0.1in}
Computations were carried out on the Blue Gene/P of the Indian Lattice
Gauge Theory Initiative, and on the Gaggle cluster of the Department 
of Theoretical Physics, TIFR. 
We thank A. Salve and K. Ghadiali for technical support. 
M. P. acknowledges support from the Austrian Science Fund FWF:I1313-N27 and the Deutsche and
Forschungsgemeinschaft Grant No. SFB/TRR 55. R.L. is supported in part by NSERC of Canada.
We are grateful to the MILC collaboration and in particular 
to S. Gottlieb for providing us with the HISQ lattices.


\vspace*{-0.1in}
\begin{thebibliography}{99}
\bibitem{PDG} C. Patrignani {\it et al.} [Particle Data Group], Chin. Phys. C, 40, 100001 (2016). 
%\cite{Aad:2014laa}
\bibitem{Aad:2014laa} 
  G.~Aad {\it et al.} [ATLAS Collaboration],
  %``Observation of an Excited $B_c^\pm$ Meson State with the ATLAS Detector,''
  Phys.\ Rev.\ Lett.\  {\bf 113}, no. 21, 212004 (2014).



\bibitem{q1}W. Kwong and J. Rosner, Phys. Rev. D 44, 212 (1991).
\bibitem{q2}E. Eichten and C. Quigg, Phys. Rev. D 49, 5845 (1994).

%\cite{Kiselev:1994rc}
\bibitem{Kiselev:1994rc} 
  V.~V.~Kiselev {\it et al.},
  %``B(c) spectroscopy,''
  Phys.\ Rev.\ D {\bf 51}, 3613 (1995).
  %%CITATION = doi:10.1103/PhysRevD.51.3613;%%
  %156 citations counted in INSPIRE as of 31 Oct 2016

%\cite{Ebert:2002pp}
\bibitem{Ebert:2002pp} 
  D.~Ebert {\it et al.},
  %``Properties of heavy quarkonia and $B_c$ mesons in the relativistic quark model,''
  Phys.\ Rev.\ D {\bf 67}, 014027 (2003).
  %%CITATION = doi:10.1103/PhysRevD.67.014027;%%
  %292 citations counted in INSPIRE as of 31 Oct 2016

%\cite{Godfrey:2004ya}
\bibitem{Godfrey:2004ya} 
  S.~Godfrey,
  %``Spectroscopy of $B_c$ mesons in the relativized quark model,''
  Phys.\ Rev.\ D {\bf 70}, 054017 (2004).
  %%CITATION = doi:10.1103/PhysRevD.70.054017;%%
  %104 citations counted in INSPIRE as of 31 Oct 2016

%\cite{Gregory:2009hq}
\bibitem{Gregory:2009hq} 
  E.~B.~Gregory {\it et al.} [HPQCD],
  %``A Prediction of the B*(c) mass in full lattice QCD,''
  Phys.\ Rev.\ Lett.\  {\bf 104}, 022001 (2010).
  %%CITATION = doi:10.1103/PhysRevLett.104.022001;%%
  %40 citations counted in INSPIRE as of 31 Oct 2016

%\cite{Dowdall:2012ab}
\bibitem{Dowdall:2012ab} 
  R.~J.~Dowdall {\it et al.} [HPQCD],
  %``Precise heavy-light meson masses and hyperfine splittings from lattice QCD including charm quarks in the sea,''
  Phys.\ Rev.\ D {\bf 86}, 094510 (2012).
  %%CITATION = doi:10.1103/PhysRevD.86.094510;%%
  %60 citations counted in INSPIRE as of 31 Oct 2016

%\cite{Wurtz:2015mqa}
\bibitem{Wurtz:2015mqa} 
  M.~Wurtz, R.~Lewis and R.~M.~Woloshyn,
  %``Free-form smearing for bottomonium and B meson spectroscopy,''
  Phys.\ Rev.\ D {\bf 92}, no. 5, 054504 (2015).
  %%CITATION = doi:10.1103/PhysRevD.92.054504;%%
  %3 citations counted in INSPIRE as of 31 Oct 2016

\bibitem{Francis:2016} A. Francis {\it et al.}, PoS LATTICE 2016, 133 (2016).

%\cite{Brown:2014ena}
\bibitem{Brown:2014ena} 
  Z.~S.~Brown {\it et al.},
  %``Charmed bottom baryon spectroscopy from lattice QCD,''
  Phys.\ Rev.\ D {\bf 90}, no. 9, 094507 (2014).

%\cite{Bazavov:2012xda}
\bibitem{Bazavov:2012xda} 
  A.~Bazavov {\it et al.} [MILC Collaboration],
  %``Lattice QCD ensembles with four flavors of highly improved staggered quarks,''
  Phys.\ Rev.\ D {\bf 87}, no. 5, 054505 (2013).

%\cite{Basak:2012py}
\bibitem{Basak:2012py} 
  S.~Basak {\it et al.}, 
  %``Charm and strange hadron spectra from overlap fermions on HISQ gauge configurations,''
  PoS LATTICE {\bf 2012}, 141 (2012).
  %%CITATION = ARXIV:1211.6277;%%
  %17 citations counted in INSPIRE as of 31 Oct 2016

%\cite{Basak:2013oya}
\bibitem{Basak:2013oya} 
  S.~Basak {\it et al.},
  %``Hadron spectra from overlap fermions on HISQ gauge configurations,''
  PoS LATTICE {\bf 2013}, 243 (2014).
  %%CITATION = ARXIV:1312.3050;%%
  %6 citations counted in INSPIRE as of 31 Oct 2016

%\cite{Lepage:1992tx}
\bibitem{Lepage:1992tx} 
  G.~P.~Lepage {\it et al.},
  %``Improved nonrelativistic QCD for heavy quark physics,''
  Phys.\ Rev.\ D {\bf 46}, 4052 (1992).
  %%CITATION = doi:10.1103/PhysRevD.46.4052;%%
  %555 citations counted in INSPIRE as of 31 Oct 2016

%\cite{Lewis:2008fu}
\bibitem{Lewis:2008fu} 
  R.~Lewis and R.~M.~Woloshyn,
  %``Bottom baryons from a dynamical lattice QCD simulation,''
  Phys.\ Rev.\ D {\bf 79}, 014502 (2009).
  %%CITATION = doi:10.1103/PhysRevD.79.014502;%%
  %70 citations counted in INSPIRE as of 31 Oct 2016

%\cite{Dowdall:2011wh}
\bibitem{Dowdall:2011wh} 
  R.~J.~Dowdall {\it et al.} [HPQCD Collaboration],
  %``The Upsilon spectrum and the determination of the lattice spacing from lattice QCD including charm quarks in the sea,''
  Phys.\ Rev.\ D {\bf 85}, 054509 (2012).
  %86 citations counted in INSPIRE as of 11 Nov 2016

%\cite{Levkova:2010ft}
\bibitem{Levkova:2010ft} 
  L.~Levkova and C.~DeTar,
  %``Charm annihilation effects on the hyperfine splitting in charmonium,''
  Phys.\ Rev.\ D {\bf 83}, 074504 (2011).

%\cite{Roberts:2007ni}
\bibitem{Roberts:2007ni} 
  W.~Roberts and M.~Pervin,
  %``Heavy baryons in a quark model,''
  Int.\ J.\ Mod.\ Phys.\ A {\bf 23}, 2817 (2008).
\end{thebibliography}
\end{document}